\documentclass[11pt]{article}
\usepackage{xspace}
\usepackage{graphicx}
\usepackage{booktabs}
\usepackage{longtable}
\usepackage{tabularx}
\usepackage{array}
\usepackage{caption}
\usepackage{placeins}
\usepackage{amsmath,amssymb}
\usepackage{microtype}
\usepackage{authblk}
\usepackage[hidelinks]{hyperref}
\hypersetup{
    pdftitle={diptych}, 
    pdflang={en}
}
\newcommand{\Description}[1]{}

\newcommand{\descTeaser}{A four-panel overview of the workflow. Panel one,
Compare: two tracks, A and B, each shown as a waveform split into labelled
sections (Intro, Verse, Chorus, Outro) with Loudness, Timbre and Rhythm
icons, over a note that the tool covers eight categories and fifty-nine
features. Panel two, Whole-song compare and AI interpretation: a verdict
card stating that Track A is more expressive than Track B, beside paired
values for the two tracks (120 versus 110 bpm, minus 13 versus minus 15
decibels, D major versus C minor, chest versus chest voice). Panel three,
User-directed segment inspection: a segment of Track A from 2:10 to 2:36
paired with a segment of Track B from 2:53 to 3:08, each with per-feature
values. Panel four, Musical judgment: a person listening on headphones,
with the steps Listen, Compare, Explore, Decide, noting that the user stays
in control of the final judgment.}

\newcommand{\descSystemOverview}{A left-to-right block diagram of the
pipeline. The user uploads two tracks, A and B. A per-track feature
extraction engine loads and resamples the audio, runs structural
segmentation, and computes fifty-nine signal-processing features grouped
into eight categories: pitch and melody, harmony and tonality, rhythm and
timing, dynamics and loudness, timbre and tone colour, emotion and
expression, vocal quality, and production and mix. This produces one
feature JSON per track holding whole-song and per-segment values. A
per-pair comparison and interpretation engine then aligns features and
segments, runs a deterministic diff, assembles a feature glossary and
numeric diff, and passes the result to a large language model served
through vLLM. The output drives the interface: a feature table, a
whole-song and segment timeline, and an AI verdict card. A return arrow
runs from the interface back to the user.}

\newcommand{\descUI}{A screenshot of the interface with three numbered
callouts. At the top, two panels, Your song (Song A) and Reference song
(Song B), each with a Replace control and nine segments. Callout one,
structure-aware segment selection: each track's timeline is divided into
coloured sections (Intro, Verse, Chorus, Bridge, Outro, Solo), and the user
pairs passages that need not share timestamps; here a segment of Song A
from 0:18 to 0:34 is paired with a segment of Song B from 0:00 to 0:22.
Callout two, AI-powered interpretation: a verdict card for Dynamics and
Loudness gives a short written comparison of the two selected segments.
Callout three, feature-level comparison: a table lists features such as
integrated loudness, dynamic range and RMS energy envelope, showing each
track's value on a shared scale and a signed difference. Two waveform
players at the bottom allow A and B playback.}

\newcommand{\descExperiment}{A two-lane flow diagram of the study. Top
lane, the musician group of twelve: a background questionnaire, a tutorial
and practice task, unaided A/B listening (recording differences, time
locations and possible next steps), a system-aided review (whole-track
comparison, segment-level evidence and AI interpretation), a review-and-
revise step (retain, revise, withdraw, or add new differences), and an
effectiveness and usability questionnaire (helpfulness, trust, SUS, and UX
feedback). Bottom lane, the expert group of four: a background
questionnaire, blind A/B listening, source-blinded claim rating, and expert
ratings. The musicians' judgments feed a shared pool of anonymized claim
units, which the experts rate. Both lanes feed a final analysis of system
validity, usability and UX, and added value after system use.}

\newcommand{\descInterp}{A horizontal stacked bar chart of ten items,
twelve participants, on a five-point agreement scale from strongly disagree
to strongly agree, with counts printed in each bar segment. Agreement was
high for noticing differences that had been missed, locating where
differences occurred, and being clearer on what to do next. Support was
weaker for the evidence being sufficient to verify the system's statements
and for separating fact from interpretation. The two reverse-worded items,
overconfidence when uncertain and being pushed toward the reference, drew
mostly disagreement or neutral responses.}

\newcommand{\descUsability}{A horizontal stacked bar chart of five items,
twelve participants, on the same five-point agreement scale, with counts
printed in each segment. All participants agreed that playback and track
switching were convenient and that the information load was manageable.
Locating the referenced segment and relating the charts to what was heard
were rated slightly lower, and inspecting the evidence behind an insight
drew the most reservations.}

\newcommand{\descHelpTrust}{Two bar charts, twelve participants each. Left,
overall helpfulness on a four-point scale: none very unhelpful, one
somewhat unhelpful, four somewhat helpful, and seven very helpful. Right,
overall trust in the results: none with no trust, two trusting slightly,
two trusting moderately, and eight trusting very much.}

\newcommand{\descSUS}{Two panels. Left, a box plot of overall System
Usability Scale scores across twelve participants, with individual points, a
mean marker, and a dashed line at the benchmark of sixty-eight; the
distribution sits above the benchmark. Right, a horizontal bar chart of the
ten SUS items, each reverse-scored so that higher is more usable on a zero
to four scale, with ninety-five percent confidence interval error bars.
Item means run from roughly 2.5 to 3.4, with easy to use and learn to use
quickly among the highest.}

\AtBeginDocument{%
  }

\begin{document}

\newcommand{\oursystem}{\textsc{DipTych}\xspace}
\newcolumntype{P}[1]{>{\raggedright\arraybackslash\hsize=#1\hsize}X}
\newcolumntype{Q}[1]{>{\hsize=#1\hsize\arraybackslash}X}
\newcolumntype{K}{>{\centering\arraybackslash}p{0.45cm}}
\newcolumntype{F}{>{\raggedright\arraybackslash}p{3.2cm}}
\newcolumntype{L}{>{\raggedright\arraybackslash}p{3.0cm}}
\newcolumntype{Y}{>{\raggedright\arraybackslash}X}
\setlength{\tabcolsep}{4pt}
\renewcommand{\arraystretch}{1.0}
\renewcommand{\topfraction}{0.95}
\renewcommand{\bottomfraction}{0.95}
\renewcommand{\textfraction}{0.05}
\renewcommand{\floatpagefraction}{0.75}
\setcounter{topnumber}{3}
\setcounter{bottomnumber}{3}
\setcounter{totalnumber}{5}
\setlength{\floatsep}{8pt plus 2pt minus 2pt}
\setlength{\textfloatsep}{8pt plus 2pt minus 2pt}
\setlength{\intextsep}{8pt plus 2pt minus 2pt}
\captionsetup[table]{font=small,skip=6pt}
\AtBeginEnvironment{table}{\small}
\newcommand{\category}[1]{%
  \midrule
  \multicolumn{4}{l}{\textbf{#1}} \\
  \midrule}

\newcommand{\feat}[1]{\texttt{#1}}
\title{Diptych: Scoped, AI-Interpreted Comparison for Reference Listening in Music Production}

\author[1,2]{Chongjun Zhong\thanks{Corresponding author: \href{mailto:zhong_chongjun@zju.edu.cn}{zhong\_chongjun@zju.edu.cn}. ORCID: \href{https://orcid.org/0000-0002-8671-050X}{0000-0002-8671-050X}}}
\author[2]{Abhinaba Roy\thanks{\href{mailto:abhinaba_roy@sutd.edu.sg}{abhinaba\_roy@sutd.edu.sg}}}
\author[2]{Archishman Ghosh\thanks{\href{mailto:archishman_ghosh@mymail.sutd.edu.sg}{archishman\_ghosh@mymail.sutd.edu.sg}}}
\author[1]{Kejun Zhang\thanks{\href{mailto:zhangkejun@zju.edu.cn}{zhangkejun@zju.edu.cn}}}
\author[2]{Dorien Herremans\thanks{\href{mailto:dorien_herremans@sutd.edu.sg}{dorien\_herremans@sutd.edu.sg}}}
\affil[1]{Zhejiang University, Hangzhou, Zhejiang, China}
\affil[2]{Singapore University of Technology and Design, Singapore}
\date{}

%

\maketitle

\begin{abstract}
Reference listening is a common strategy in music production, but current comparison tools often obscure a key human judgment: deciding what should be compared. We present \oursystem{}, an AI-assisted system that lets users define comparison scope across whole tracks or independently selected segments, while inspecting structured audio features and scope-specific AI interpretations. We evaluated \oursystem{} in a within-participants study with 12 musicians, complemented by source-blinded ratings from four expert listeners. Participants used the system to surface additional differences, nine of ten of which received at least partial expert support, and reported good usability and greater clarity about possible next steps. These findings suggest that AI support for creative comparison should prioritize user-defined scope, inspectable evidence, and actionable guidance, while avoiding authoritative judgments that exceed what the evidence can support.
\end{abstract}



\begin{figure}[t]
 \centering
 \includegraphics[width=\linewidth]{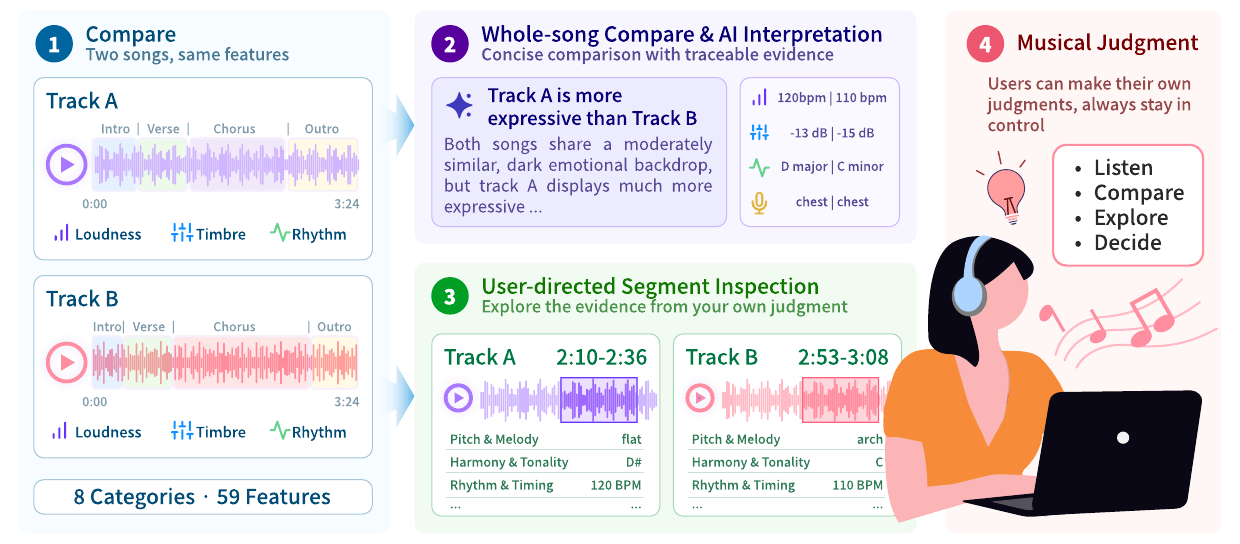}
  \caption{\oursystem{} supports reference listening across multiple scales: users compare whole tracks, inspect AI-interpreted feature differences, and select specific segments for closer examination while retaining control over what to compare and how to judge the result.}
  \Description{\descTeaser}
  \label{fig:teaser}
\end{figure}

\section{Introduction}
Music producers and mixing engineers rarely work in isolation from the music they admire. When shaping a new track, they pull up a commercial recording as a reference, and move back and forth between it and their own work in progress. Existing songs have also been used as examples for communicating musical intent in HCI systems for AI-assisted music creation~\cite{frid2020music}. This practice, reference listening, is a standard part of professional mixing and mastering workflows~\cite{vanka2024referencesongs}: an engineer plays a section, plays the same moment in the reference, notices the chorus does not open up as much, and decides what to change.

Before making such a judgment, however, the engineer has already decided which two passages are the right ones to compare. Two songs rarely line up: they differ in tempo, arrangement, length, and structure~\cite{serra2010cover}, so a chorus that lands two minutes into one track may sit thirty seconds into the other, run for a different duration, or carry a different structural weight. Deciding which regions of two tracks are worth comparing, and at what temporal scope, is not a step that precedes reference listening. It is part of the listening itself.

This decision is also hard to hold in memory. Comparing two choruses means keeping one in mind while playing the other, matching a fading impression against what is currently being heard. Auditory two-stimulus comparison becomes harder once a stimulus can no longer be checked against a fresh sensory trace and must instead rely on a more fragile working-memory representation~\cite{nees2016auditorymemory}. As sections or comparison points multiply, so does this memory burden: much of the difficulty lies not in hearing a difference, but in holding the two passages still long enough to compare them.

A second difficulty is articulation: musicians and production engineers often hear more than they can express. Recording engineers may struggle to locate, name, or explain a problem in a mix, and developing the vocabulary to do so is itself part of professional training~\cite{porcello2004speaking}; building shared, communicable vocabularies for sound qualities remains an open design problem~\cite{carron2017speaking}. Reference listening therefore also requires turning a vague impression such as ``something feels wrong here'' into a specific, checkable judgment such as ``the reference has more space around the vocal in the second chorus.''

Common tools that support reference comparison work at the level of the whole track, reporting differences over the full recording~\cite{deman2019imp,deman2017tenyears}. This is useful for broad spectral balance, but gives engineers no way to say ``hold my chorus against their chorus,'' or against whichever passage they judge to be the fair comparison. Deciding what maps onto what is therefore exactly the decision these systems make for them.

Recent AI systems make this gap sharper. A capable model can produce a fluent, technically correct account of the differences between two pieces of audio~\cite{mixassist2025}, but the harder problem is deciding where to look in the first place. Even in adjacent domains, AI-generated feedback has been shown to trail human evaluators specifically on prioritizing what matters and staying accurate while remaining comparable in fluency and structure~\cite{steiss2024comparing}. Related HCI work shows that convincing LLM explanations can increase reliance even when the underlying response is incorrect~\cite{kim2025fostering}. A system can generate an accurate account of a difference the user never cared about, or one pitched at the wrong scope. Correctness and usefulness come apart here: an AI verdict about the wrong comparison is not rescued by being accurate. There is also a reason usefulness is the right test. A producer comparing two tracks is trying to decide what to do next: change something, try something, or leave it alone. A difference that is real but says nothing about what to do about it does not help much. A comparison earns its place when it moves the user toward a decision, not merely when it reports something true.\par

We build \oursystem{} around this observation. The user decides what to compare, and the AI interprets structured audio analysis within that chosen scope. The system offers two comparison modes. Whole Song mode compares complete tracks, while Selected Segments mode lets the user select a region from each track independently, without requiring a shared timestamp, duration, or structural label. The musical structure of each track is laid out on a timeline, so the user can find sections quickly rather than holding the layout of both tracks in memory. In both modes the user sees a set of structured audio features alongside an AI-generated verdict. The verdict is presented as an interpretation rather than an objective judgment of musical quality, and users remain free to accept, revise, or discard it.

Our goal is not only to show that this one system works. We want to surface design principles for AI-assisted music comparison more broadly, so that the wider community building these tools has something to reuse. Reference listening is one case of a broader pattern in which people use examples to support creative work~\cite{herring2009getting}, here with an AI helping read the difference. Related patterns appear in video editing, where creators compare rough-cut alternatives against source footage and each other~\cite{huh2025videodiff}, and in writing, where learners revise drafts against model texts~\cite{wu2023modeltexts}. Building \oursystem{} forced a set of choices that any such tool has to make: who decides what gets compared, the person or the system; how closely the AI's interpretation should be tied to the scope the person chose; how the raw evidence and the written verdict should sit next to each other, so the person can check one against the other; and which features are worth putting in front of the person in the first place, out of the many that could be measured. We treat these as design questions worth answering in general, and use our system as a way to study them rather than as the end point.

We study this design with 12 musicians in a single-session study. Participants first make unaided judgments on a fixed target-reference pair, then use the full system and revisit those judgments by retaining, revising, withdrawing, or adding claims. They also report on the system's effectiveness, helpfulness, trust, usability, and overall experience. We recruit four expert listeners as a perceptual reference. They listen to the same tracks by ear, without the system, and then judge the claims the participants made, both before and after using the system. Because the experts work without the interface, they tell us how much it actually helped the participants, and how much of what participants claimed is genuinely audible. Three questions organize the work: 

\begin{itemize}
\item RQ1: how do musicians' comparative judgments change after using the system?
\item RQ2: how do musicians adopt, revise, or qualify the system's comparison interpretations in their judgments? 
\item RQ3: to what extent are those interpretations supported by expert listening, and how do users rate the system's usefulness and usability?
\end{itemize}

This paper makes three contributions. First, we present an interaction design that keeps comparison scope under user control, across both whole tracks and independently selected segments. Second, we provide an empirical account of how musicians' comparative judgments change when they can set that scope and inspect structured analysis alongside AI interpretation. Third, an independent expert panel that checks the system's comparisons against careful listening, from which we draw design principles for AI-assisted music comparison. The central one is that being correct is not enough: a comparison can be accurate and still tell the user very little, so it also has to fit the scope, be specific, and point toward a decision the user can act on. Although we study music, the same split between choosing what to compare and reading the difference recurs in other reference-based creative work, from video editing to document revision, wherever a person holds their work against an example.

\section{Related Work}

Our work is related to three primary areas: (1) AI tools that take part in making music, (2) tools that help people compare while they produce it, and (3) audio analysis that turns a recording into interpretable numbers.

\subsection{AI-Assisted and Co-Creative Music Tools}

Recent years have seen a number of generative models that create a full track from a text prompt~\cite{agostinelli2023musiclm,melechovsky2024mustango,liu2025jam,bhandari2025text2midi}, as well as models that assist with edits to a work in progress~\cite{zhang2024musicmagus,melechovsky2025sonicmaster,roy2026text2midi}. HCI has also explored example-based music creation~\cite{frid2020music}, raising a longstanding mixed-initiative question~\cite{deterding2017mixed}: how control should be split between person and system. Broader guidelines for AI-infused interfaces converge on the same answer regardless of domain: keep the person able to inspect, correct, and dismiss the system's output rather than accept it wholesale~\cite{amershi2019guidelines}. In music specifically, steering controls and musician-centered design similarly highlight the importance of user agency~\cite{louie2020novice,krol2025exploring}.
 
A second line turns audio into language. Captioning models, increasingly built on general-purpose audio-language representations~\cite{li2024mert,wu2023clap}, describe a clip in free text~\cite{doh2023lp,lanzendorfer2025bootstrapping}, and some models read a clip's emotion~\cite{kang2026towards,liu2024leveraging}. These answer one question: what does this sound like. They decide what to say about a clip, and at what level of detail.

In both cases, the system typically holds the initiative: it generates content or decides what to describe. Our work instead keeps the comparison choice with the user, while the system responds within that scope. Prior HCI work on AI-assisted musical improvisation likewise highlights the value of exposing system state to musicians~\cite{mccormack2019silent}. The judgment remains with the user, consistent with work emphasizing that explanations should support people in checking, questioning, or rejecting system output~\cite{miller2019explanation}.
 
\subsection{Comparison and Reference-Conditioned Support in Music Production}
Producers routinely compare a work in progress against a finished commercial track, and a substantial body of intelligent-music-production research has grown up around this practice~\cite{deman2019imp,deman2017tenyears}. One branch of this work targets the comparison itself: plugins and analysis tools report the whole-track frequency balance, loudness, and stereo image of a mix against a reference.\footnote{iZotope Ozone, Logic Pro.} These tools compare whole tracks, and most commonly work with EQ and tonal balance. This is a reasonable default for broad spectral balance, but it has a shortcoming: if a chorus arrives two minutes into one track and thirty seconds into the other, the tool never holds one chorus against the other, because it has no way to know the user wanted those two parts compared specifically.

A second, more active branch does not report a comparison to the user at all; instead, it uses the reference to automatically change the target. Differentiable-mixing-console systems predict per-track effect parameters end to end from a set of raw stems~\cite{steinmetz2021automatic}, style-transfer systems learn to reproduce a reference recording's spectral and dynamic signature on a new signal~\cite{steinmetz2022style}, and encoder-based systems disentangle audio-effect style from musical content so it can be transplanted from a reference song onto a raw mix~\cite{koo2023mixingstyle,martinezramirez2022automatic}. The most complete instance of this line, Diff-MST, predicts full mixing-console parameters for up to twenty tracks directly from a reference song, offering manual adjustment only after an initial mix has already been generated~\cite{vanka2024diffmst}. HCI work has similarly used existing songs as examples to condition interactive AI music generation~\cite{frid2020music}. A related, more conversational line trains audio-language models to give mixing advice in dialogue with the producer~\cite{mixassist2025}, moving beyond a fixed set of numeric controls but still deciding, on the model's own terms, what to talk about and when.
 
What distinguishes our work is where the decision about scope sits. Whole-track tools do not expose local comparison scope, while transfer and generative systems use the reference to produce an output rather than present a comparison. In our design, users choose the regions to compare, even when they differ in start time, length, or structural label, and the system returns an interpretation for them to inspect and act on. Related HCI work on audio editing similarly keeps temporal selection user-directed while allowing automation to refine it~\cite{shi2018loopmaker}. This follows a design principle established in visual-comparison research more broadly: effective interfaces for comparing complex objects juxtapose the objects being compared and layer explicit, inspectable encodings of their differences on top, rather than collapsing the comparison into a single merged or automatically resolved output~\cite{gleicher2011visual}. Our region-scoped feature table and AI verdict are, in this sense, an audio instance of a design pattern HCI has mostly studied through visual and textual objects.
 
\subsection{A Structured Vocabulary for What Gets Compared}
\label{sec:feature-foundations}
 
A single similarity score collapses everything a listener might notice into one number, and in doing so it hides what actually differs and where. If a target track differs from its reference in loudness but not in tonal balance, a scalar distance says only that something differs, not what a producer should go and listen for. Recent work confirms the shape of this problem directly at the representation level: MERIT~\cite{roy2026merit} shows that standard audio embeddings entangle melody, rhythm, and timbre into a single vector, and demonstrates that training separate, disentangled representations for each of these three factors yields similarity scores that better track what listeners actually judge two recordings to share along a given dimension~\cite{roy2026merit}. We start from the same observation, that a monolithic similarity score hides more than it reveals, and go further in two respects: where MERIT recovers three perceptual factors as learned embeddings for retrieval, we expose eight production-relevant dimensions as interpretable, user-scoped comparisons, extending the same disentanglement principle from a fixed embedding space into features a producer can read, check, and act on directly. Comparison output is useful only if it is organized at the grain a producer already thinks in: not raw signal, not one collapsed number, and not even a handful of retrieval-oriented factors, but a small set of musical dimensions that map onto production decisions.
 
We organize our comparison around eight such dimensions, chosen not arbitrarily but where three independent sources of evidence converge. First, the domains mix engineers themselves reach for when describing a reference track are close to these eight: professional engineers routinely invoke tonal balance, dynamics, and spatial characteristics when explaining what a reference song should communicate~\cite{vanka2024referencesongs}, tools have been built specifically to give practitioners a structured, shared vocabulary for naming sound qualities along comparable dimensions~\cite{carron2017speaking}, and early work on autonomous mixing found it necessary to compile practical mixing-engineering literature into exactly this kind of rule set, organized by production domain rather than by raw signal feature, before any system could act on it usefully~\cite{deman2013semantic}. Second, the MIR community has long organized audio content description around a small number of facets, rhythm, harmony, timbre, and mood or emotion among them, rather than a single audio embedding, precisely because different retrieval and comparison tasks call on different musical dimensions~\cite{casey2008contentbased}. Third, work on formally representing music production knowledge has found it necessary to build explicit taxonomies of audio effects and production decisions in order to make that knowledge shareable and machine-readable at all~\cite{wilmering2013audioeffects}. Practitioner vocabulary, MIR facet organization, and production-knowledge engineering arrive, independently, at essentially the same partition of what there is to compare.
 
This convergence also motivates categories that might otherwise look like a subdivision of a broader one. Vocal quality is a case in point: rather than folding vocal character into general timbre, music-theoretic work on vocal performance argues that vocal timbre carries interpretive and affective information that generic spectral description does not capture, and needs its own descriptive vocabulary as a result~\cite{heidemann2016vocaltimbre}. We follow that distinction and keep vocal quality as its own category.
 
Within each dimension, individual features are chosen because they give a producer something specific and checkable to act on, not just a number that moves. A few example features from each category illustrate the pattern; the complete feature set and extraction pipeline are described in Section~\ref{sec:implementation}.
 
\paragraph{Pitch and melody.} An example feature in this category is vibrato rate: professional singers converge on a narrow, well-documented range of oscillation rates, so a same-melody comparison that shows the reference's vibrato as faster or slower gives a producer something concrete to react to, rather than a vague sense that something sounds different~\cite{prame1994vibrato}. Melodic contour, the up-down shape a line traces independent of its exact intervals, is a second example: it has long been treated as a distinct, describable property of a melody in its own right, which is why we track it as a comparable shape rather than only as a sequence of pitch values~\cite{adams1976melodic}.
 
\paragraph{Harmony and tonality.} An example feature here is harmonic change: representing pitch-class content on a Tonnetz-derived tonal space and tracking movement through it gives a measure that flags exactly where two progressions diverge, rather than only whether they share a key~\cite{harte2006harmonic}. Key clarity is a second example, built on the classic finding that listeners hold a graded sense of how strongly a passage centres on a given key rather than a binary in-key or out-of-key judgment, which is what a clarity score is intended to capture~\cite{krumhansl1982tracing}.
 
\paragraph{Rhythm and timing.} An example feature is swing ratio, the systematic long-short durational bias between consecutive eighth notes: ensembles converge on characteristic ratios that shift with tempo, and a producer comparing groove against a reference is often, without knowing the term, listening for exactly this~\cite{friberg2002swing}. Syncopation is a second example, quantified since the 1980s as a function of how strongly a note falls on a metrically weak position relative to the surrounding pulse, giving groove differences a score rather than only a qualitative label~\cite{longuethiggins1984rhythmic}.
 
\paragraph{Dynamics and loudness.} An example feature is integrated LUFS and loudness range, the broadcast-standard perceptual loudness measures, which track what a producer actually hears as louder or more dynamic rather than just what clips first~\cite{series2011algorithms,ebu2011loudness}. Attack transient sharpness is a second example: how quickly a sound is perceived to begin, its perceptual attack time, is known to vary with a note's rise time independently of its peak amplitude, which is why we treat onset sharpness as its own measurable quantity rather than folding it into loudness alone~\cite{gordon1987perceptual}.
 
\paragraph{Timbre and tone colour.} An example feature is the H1-H2 measure, the relative amplitude of the first two harmonics, developed specifically to quantify breathiness and glottal configuration and giving a concrete, checkable account of why one take sounds airier than another~\cite{hanson1997glottal}. Spectral centroid is a second example: it is the best-established acoustic correlate of perceived brightness, one of the first dimensions listeners reach for when distinguishing two otherwise similar tones~\cite{grey1977multidimensional}.
 
\paragraph{Emotion and expression.} An example feature is the valence-arousal circumplex, which reduces perceived affect to two axes that can be tracked and compared directly~\cite{russell1980circumplex}. A parametric tension model is a second example, capturing how perceived tension rises and falls across a passage and giving a shape a producer can compare against a reference's build and release~\cite{farbood2012tension}. Micro-timing feel, the small, systematic timing deviations that give a performance its rubato or lack of it, is a third example: detailed analyses of expressive timing in performance show these deviations are structured and characteristic of a performer rather than noise, which is why they are worth comparing rather than averaging away~\cite{repp1992diversity}.
 
\paragraph{Vocal quality.} As argued above, vocal quality needs its own vocabulary because it carries meaning that generic timbre description does not~\cite{heidemann2016vocaltimbre}. An example feature is formant configuration, a well-established way to operationalize this, from the lower formants that define vowel identity to the higher-formant clustering that gives a trained voice its characteristic carrying power~\cite{sundberg1974singing}. Jitter and shimmer, cycle-to-cycle variation in pitch and amplitude, are a second example, developed to characterize exactly the kind of voice-quality difference, roughness, strain, instability, that a spectral snapshot alone does not capture~\cite{farrus2007jitter}.
 
\paragraph{Production and mix.} An example feature is stereo width and the kick-bass low-end relationship, which the reference-conditioned mixing literature discussed in Section~2.2 treats as exactly the parameters worth matching to a reference~\cite{steinmetz2022style,vanka2024diffmst}, and which is why we treat them as dimensions worth comparing rather than acting on automatically. Reverb decay time is a second example: listener tolerance for how much artificial reverberation a mix carries, and how quickly it decays, has been shown to occupy a fairly narrow acceptable range with a measurable effect on perceived mix quality, making it a concrete, checkable point of comparison rather than a vague sense of space~\cite{deman2017reverberation}.

What none of this prior work does is put all of these dimensions in front of a user at once, for a comparison the user has scoped themselves. Systems that report structured feature differences tend to specialize: chord-recognition tools address harmony alone~\cite{akram2025chordformer}, loudness-normalization standards and automatic spectral-balance tools address dynamics and tone alone~\cite{series2011algorithms,ebu2011loudness,mockenhaupt2024automatic}, mood- and emotion-recognition models address a single affective dimension~\cite{kang2026towards,liu2024leveraging}, captioning models collapse everything back into a single free-text description~\cite{doh2023lp,lanzendorfer2025bootstrapping}, and a factor-disentangled embedding model such as MERIT returns similarity scores for retrieval rather than an inspectable account of a specific comparison~\cite{roy2026merit}. Section~\ref{sec:implementation} lists the complete feature set within each of these eight dimensions and how it is extracted; the categories themselves are what let us keep the evidence for a comparison organized the way a producer already organizes their listening, rather than as an undifferentiated feature vector or a single collapsed score.

\section{System Design and Implementation}
Reference listening asks the user to do two things. First they decide which parts of the two tracks to hold against each other, and then they read the difference between them. The introduction argued that a helpful tool should support the second step without taking over the first. We built \oursystem{} around that idea. This section describes its architecture, the comparison interface and its two modes, the AI interpretation, and how each part is implemented.
\subsection{System Overview and Architecture}

\oursystem{} keeps the choice of what to compare with the user. A meaningful comparison may hold two entire tracks against each other, but it may also involve two passages that fall at different times or last for different durations. The system therefore treats comparison scope as an explicit user choice, not something it works out on its own from the music.  

Based on this, we set two design goals. First, the system should let users define and revise the comparison scope while retaining enough structural context to navigate both tracks. Second, it should turn audio-analysis output into a useful comparative account without presenting it as a substitute for listening or as a judgment of musical quality. Structural segmentation gives the user landmarks for finding sections in each track, so they do not have to hold the layout of both tracks in their head. The user still decides which sections belong together.

\begin{figure}[t]
    \centering
    \includegraphics[width=1.0\linewidth]{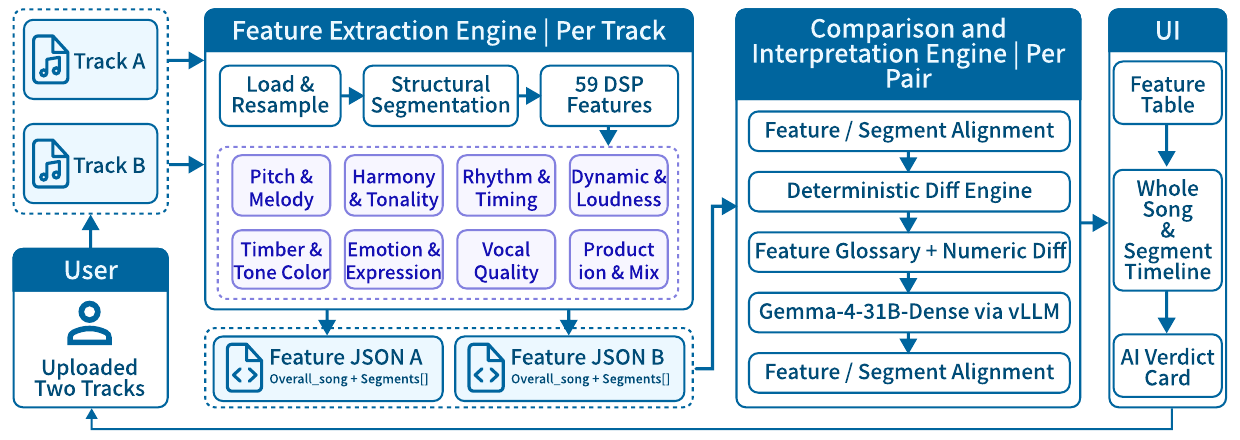}
    \caption{System architecture. The target and reference tracks are independently segmented and represented by shared audio features. Whole-track or user-selected segment scopes are passed to the comparison and interpretation engine, whose results are presented in a common comparison workspace.}
    \Description{\descSystemOverview}
    \label{fig:system-overview}
\end{figure}

Figure~\ref{fig:system-overview} shows the resulting architecture. A target track and a reference track are independently resampled, structurally segmented, and represented using a shared set of audio features. The comparison and interpretation engine derives scope-specific differences from these representations. The interface then brings together the feature tables, whole-track and segment timelines, and the AI verdict so that users can move between listening, inspecting evidence, and interpreting the current comparison.

\subsection{Comparison Interface and Interaction}
The comparison interface brings together the target and reference tracks, their timelines, a feature table, and an AI verdict card. Users can listen to the tracks and inspect their differences in two modes: Whole Song and Selected Segments. Both modes use the same feature categories, so users can move from a broad comparison to a local one without learning a different display. Each row in the table shows the target and reference values for one feature. Users can expand the categories they want to inspect.

\begin{figure}[t]
    \centering
    \includegraphics[width=1.0\textwidth]{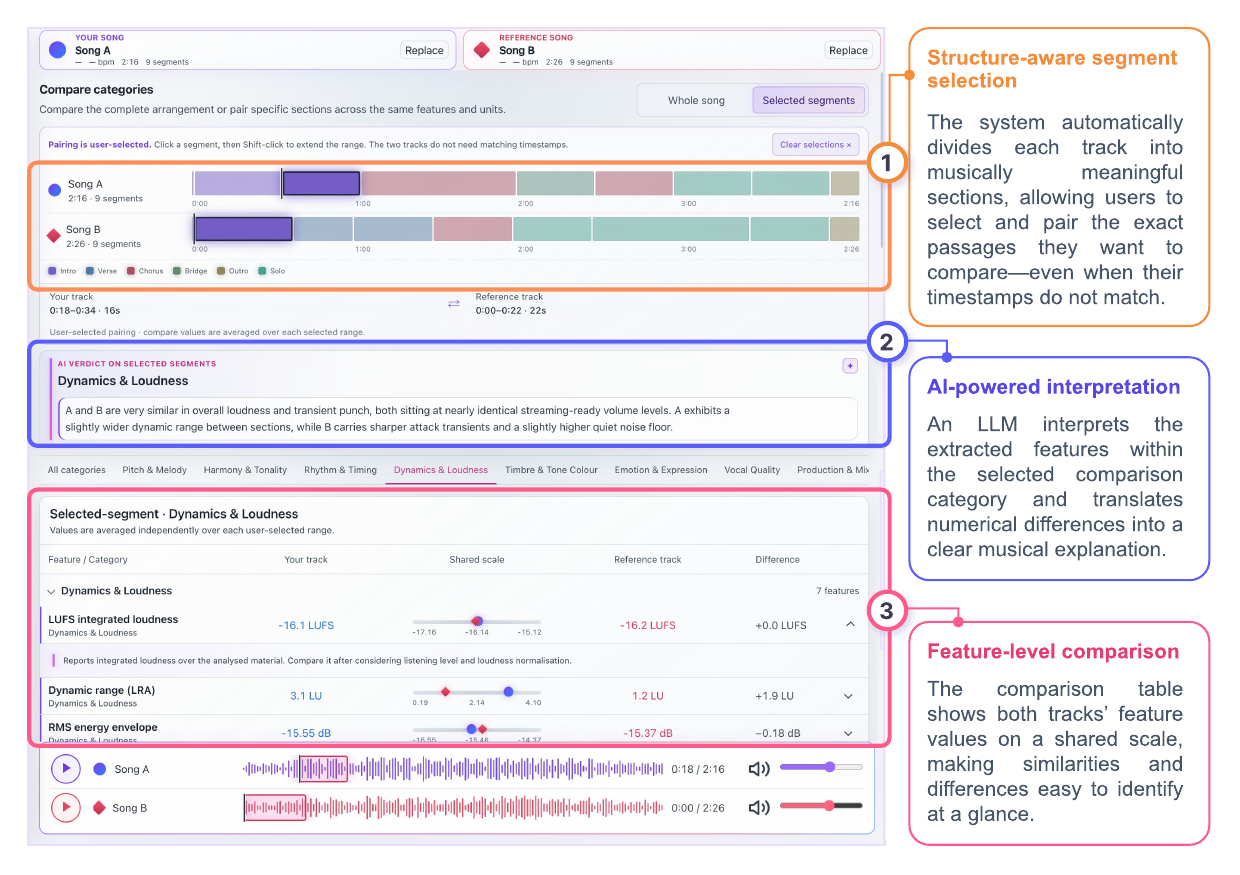}
    \caption{\oursystem’s selected-segment comparison interface, combining structure-aware segment selection, AI-generated interpretation, and feature-level comparison to help users inspect differences between independently chosen passages.}
    \Description{\descUI}
    \label{fig:placeholder}
\end{figure}

\subsubsection{Whole-Track Comparison}

Whole Song mode compares the two complete recordings. The feature table shows their overall characteristics, and the AI verdict summarizes the comparison. This mode supports questions about the tracks as a whole, such as how their loudness, rhythmic feel, or tonal balance differs. Users can inspect a category and listen again to consider how the reported difference relates to what they hear.

However, a whole-track summary can hide changes within a music track. For example, a track may be quiet overall but have a loud chorus. Thus, users can move to Selected Segments mode to examine the passages that matter to their comparison.

\subsubsection{User-Selected Segment Comparison}

The system divides each track into sections based on its musical structure and displays them on a colour-coded timeline. Labels such as Intro, Verse, Chorus, Bridge, Outro, and Solo help users locate passages.

Users choose a segment from each track independently. The two segments can have different start times, lengths, or structural labels. A user may compare two choruses, or compare a verse in one track with a chorus in the other if that better serves their listening goal. Choosing a segment in one track does not select its counterpart in the other.

Once both segments are selected, the system updates the feature comparison and generates an AI verdict for that pair. Users can inspect the results, listen to the passages, and change either selection to explore another comparison. Structural segmentation helps users find sections to compare, while the choice of which sections belong together remains theirs.

\subsection{AI Interpretation}
\label{sec:scope-conditioned-interpretation}

The AI feedback includes a summary for each feature category and an overall verdict. It describes differences in words so that users can consider several measurements together. The feature values remain available alongside the feedback for closer inspection. Putting a difference into words, next to where it occurs, gives the user a way to pin down an impression that was still vague.\par
Each interpretation concerns the current comparison. In Whole Song mode, it describes the complete recordings. In Selected Segments mode, it describes only the two chosen passages. Changing the selection produces a new interpretation for the new pair.\par
The verdict is meant to help the user decide what to do next, whether to change something, try something, or leave it alone, not just to state a true difference. Users can weigh it against the feature values and their own impressions, and they are free to accept it, revise it, or set it aside. It does not rank the recordings by musical quality or ask the target to match the reference.

\subsection{Implementation}
\label{sec:implementation}
\paragraph{Audio analysis.}
\label{sec:audio-representation}

Drawing on the music-information-retrieval, audio-analysis, and music-production literature discussed in Section~\ref{sec:feature-foundations}, we assembled 59 audio features in eight categories. The extraction engine processes each track independently using librosa. It decodes the audio and resamples it to 22.05~kHz for spectral and harmonic analysis and 8~kHz for pitch tracking. The pipeline includes YIN/pYIN pitch tracking, chroma features, spectral descriptors, mel-frequency cepstral coefficients, onset and beat tracking, and RMS and loudness statistics. Features are computed for the full track and each structural segment. One JSON file per track stores an \texttt{overall\_song} block and a \texttt{segments} array, with time boundaries attached to the segment records.


\begin{table}[!b]
\centering
\caption{Melody features: definitions, extraction library, and supporting reference.}
\label{tab:features-melody}
\begin{tabularx}{\linewidth}{@{} K F L Y @{}}
\toprule
\# & Feature & Library & What it captures \\
\midrule
1 & \texttt{Melodic contour}      & librosa~\cite{mcfee2015librosa}      & Shape and direction of the melodic line over time: rising vs.\ falling motion, tracked as continuous movement~\cite{adams1976melodic}. \\
2 & \texttt{Pitch accuracy}       & librosa~\cite{mcfee2015librosa}      & Intonation: how close notes sit to the nearest tempered pitch, in cents~\cite{dallabella2007singing}. \\
3 & \texttt{Vibrato rate}               & scipy.signal\allowbreak\ \cite{virtanen2020scipy} & Presence, rate, and extent of periodic pitch modulation~\cite{prame1994vibrato}. \\
4 & \texttt{Spectral flux}        & librosa~\cite{mcfee2015librosa}      & Frame-to-frame spectral change: how fast the sound's energy or timbre shifts~\cite{bello2005onset}. \\
5 & \texttt{Melodic interval entropy}    & librosa~\cite{mcfee2015librosa}      & Distribution of interval sizes between consecutive notes (stepwise motion vs.\ leaps)~\cite{vostroost1989ascending}. \\
6 & \texttt{pitch\_range}          & librosa~\cite{mcfee2015librosa}      & Span between lowest and highest pitch; overall tessitura~\cite{vonhippel2000redefining}. \\
7 & \texttt{Tonal stability}      & librosa~\cite{mcfee2015librosa}      & How steadily the pitch centre holds versus drifts over time~\cite{krumhansl1982tracing}. \\
8 & \texttt{Harmony pitch\_class distribution} & librosa~\cite{mcfee2015librosa}      & Histogram of which pitch classes are present (chroma-style distribution)~\cite{fujishima1999realtime}. \\
\bottomrule
\end{tabularx}
\end{table}

\begin{table}[!b]
\centering
\caption{Rhythm features: definitions, extraction library, and supporting reference.}
\label{tab:features-rhythm}
\begin{tabularx}{\linewidth}{@{} K F L Y @{}}
\toprule
\# & Feature & Library & What it captures \\
\midrule
1 & \texttt{Tempo}                    & Beat This!~\cite{foscarin2024beat}                        & Tempo in BPM~\cite{moelants2002preferred}. \\
2 & \texttt{Beat strength profile}           & Beat This! + librosa~\cite{foscarin2024beat,mcfee2015librosa} & How pronounced and driving the beat is~\cite{parncutt1994pulse}. \\
3 & \texttt{Rhythmic swing ratio}                    & librosa + Beat This!~\cite{mcfee2015librosa,foscarin2024beat} & Swing vs.\ straight feel (swing ratio)~\cite{friberg2002swing}. \\
4 & \texttt{Note onset timing deviation} & librosa + Beat This!~\cite{mcfee2015librosa,foscarin2024beat} & Where notes land relative to the grid: ahead of or behind the beat~\cite{iyer2002embodied}. \\
5 & \texttt{Rhythmic syncopation score}              & librosa + Beat This!~\cite{mcfee2015librosa,foscarin2024beat} & Degree of off-beat emphasis~\cite{longuethiggins1984rhythmic}. \\
6 & \texttt{Climax moment position}        & librosa~\cite{mcfee2015librosa}                            & How and where energy builds toward a peak~\cite{farbood2012tension}. \\
7 & \texttt{IOI regularity}          & librosa~\cite{mcfee2015librosa}                            & Consistency of inter-onset intervals: timing steadiness~\cite{largepalmer2002perceiving}. \\
8 & \texttt{Rhythmic density}        & librosa + Beat This!~\cite{mcfee2015librosa,foscarin2024beat} & Onsets or notes per unit time~\cite{senn2017rhythmic}. \\
\bottomrule
\end{tabularx}
\end{table}

\begin{table}[!b]
\centering
\caption{Emotion and expression features: definitions, extraction library, and supporting reference.}
\label{tab:features-emotion}
\begin{tabularx}{\linewidth}{@{} K F L Y @{}}
\toprule
\# & Feature & Library & What it captures \\
\midrule
1 & \texttt{Valence-arousal} & music2emo~\cite{kang2026towards} & Valence/arousal point and its emotion quadrant~\cite{russell1980circumplex}. \\
2 & \texttt{Micro-timing expressive feel}    & librosa~\cite{mcfee2015librosa}  & Micro-deviations in timing: tight vs.\ loose feel~\cite{iyer2002embodied}. \\
3 & \texttt{Phrase-level dynamics} & librosa~\cite{mcfee2015librosa}  & Loudness shape of phrases (arch, crescendo, flat, etc.)~\cite{friberg2006kth}. \\
4 & \texttt{Legato/staccato balance} & librosa~\cite{mcfee2015librosa}  & Articulation: connected vs.\ detached notes~\cite{bresinbattel2000articulation}. \\
5 & \texttt{rubato}           & librosa~\cite{mcfee2015librosa}  & Tempo elasticity vs.\ strict quantization~\cite{repp1992diversity}. \\
6 & \texttt{Tension-release curve} & librosa~\cite{mcfee2015librosa}  & Tension arc over time: buildup, peaks, and release~\cite{farbood2012tension}. \\
\bottomrule
\end{tabularx}
\end{table}

\begin{table}[!b]
\centering
\caption{Harmony features: definitions, extraction library, and supporting reference.}
\label{tab:features-harmony}
\begin{tabularx}{\linewidth}{@{} K F L Y @{}}
\toprule
\# & Feature & Library & What it captures \\
\midrule
1 & \texttt{Key and mode}       & librosa~\cite{mcfee2015librosa} & Estimated key and major/minor mode~\cite{krumhansl1982tracing}. \\
2 & \texttt{Chord progression}   & librosa~\cite{mcfee2015librosa} & Sequence and distribution of detected chords~\cite{fujishima1999realtime}. \\
3 & \texttt{Chroma profile}      & librosa~\cite{mcfee2015librosa} & Average energy per pitch class across the 12-bin chroma~\cite{fujishima1999realtime}. \\
4 & \texttt{Harmonic change}     & librosa~\cite{mcfee2015librosa} & Rate of harmonic movement (harmonic flux)~\cite{harte2006harmonic}. \\
5 & \texttt{Tonnetz centroid}    & librosa~\cite{mcfee2015librosa} & Position in tonal-centroid (Tonnetz) space: the track's harmonic ``location''~\cite{harte2006harmonic}. \\
6 & \texttt{Key clarity}         & librosa~\cite{mcfee2015librosa} & How strongly a single key is implied vs.\ ambiguous or atonal~\cite{krumhansl1982tracing}. \\
7 & \texttt{Mode majorness}      & librosa~\cite{mcfee2015librosa} & Degree of major- vs.\ minor-ness on a continuum~\cite{krumhansl1982tracing}. \\
8 & \texttt{Harmonic complexity} & librosa~\cite{mcfee2015librosa} & Chromaticism: how far the harmony sits from simple diatonic writing~\cite{lerdahl2001tonal}. \\
\bottomrule
\end{tabularx}
\end{table}

\begin{table}[!b]
\centering
\caption{Dynamics and loudness features: definitions, extraction library, and supporting reference.}
\label{tab:features-dynamics}
\begin{tabularx}{\linewidth}{@{} K F L Y @{}}
\toprule
\# & Feature & Library & What it captures \\
\midrule
1 & \texttt{LUFS integrated loudness}              & pyloudnorm\allowbreak\ \cite{steinmetz2021pyloudnorm} & Integrated loudness against the streaming target~\cite{series2011algorithms,ebu2011loudness}. \\
2 & \texttt{Dynamic range(LRA)}            & pyloudnorm\allowbreak\ \cite{steinmetz2021pyloudnorm} & Loudness range (LRA) across the track~\cite{ebu2011loudness}. \\
3 & \texttt{RMS energy envelope}          & librosa~\cite{mcfee2015librosa}           & RMS energy over time and its statistical distribution~\cite{series2011algorithms}. \\
4 & \texttt{Crest factor}                  & numpy~\cite{harris2020numpy}              & Peak-to-RMS ratio: transient health vs.\ squashing~\cite{gordon1987perceptual}. \\
5 & \texttt{Sectional loudness contrast}  & pyloudnorm\allowbreak\ \cite{steinmetz2021pyloudnorm} & Loudness gap between the loudest and quietest sections, at song level~\cite{ebu2011loudness}. \\
6 & \texttt{Attack transient sharpness}   & librosa~\cite{mcfee2015librosa}           & How sharp vs.\ soft the attacks are~\cite{gordon1987perceptual}. \\
7 & \texttt{Quiet floor level}            & pyloudnorm\allowbreak\ \cite{steinmetz2021pyloudnorm} & Loudness of the quiet floor: intimacy vs.\ always-loud~\cite{series2011algorithms}. \\
\bottomrule
\end{tabularx}
\end{table}

\begin{table}[!b]
\centering
\caption{Production and mix features: definitions, extraction library, and supporting reference.}
\label{tab:features-production}
\begin{tabularx}{\linewidth}{@{} K F L Y @{}}
\toprule
\# & Feature & Library & What it captures \\
\midrule
1 & \texttt{Stereo width}              & librosa~\cite{mcfee2015librosa}                            & Width of the stereo image, from mono to wide~\cite{deman2019imp}. \\
2 & \texttt{Frequency-band balance}   & librosa~\cite{mcfee2015librosa}                            & Tonal balance across sub/bass/low-mid/mid/air bands~\cite{mockenhaupt2024automatic}. \\
3 & \texttt{Reverb decay time (RT60)}        & librosa + scipy.signal\allowbreak\ \cite{mcfee2015librosa,virtanen2020scipy} & Reverb tail length (RT60): dryness vs.\ wetness/muddiness~\cite{deman2017reverberation}. \\
4 & \texttt{Compression artifacts}     & librosa + scipy.signal\allowbreak\ \cite{mcfee2015librosa,virtanen2020scipy} & Detection of pumping / over-compression~\cite{deman2019imp}. \\
5 & \texttt{Kick-bass low-end relationship}   & librosa~\cite{mcfee2015librosa}                            & Low-end masking vs.\ separation between kick and bass~\cite{steinmetz2022style,vanka2024diffmst}. \\
6 & \texttt{True peak level}          & pyloudnorm\allowbreak\ \cite{steinmetz2021pyloudnorm}                  & Inter-sample true peak; clipping risk~\cite{series2011algorithms}. \\
7 & \texttt{Noise floor \& room tone}   & librosa + scipy.signal\allowbreak\ \cite{mcfee2015librosa,virtanen2020scipy} & Level of background noise / room tone~\cite{deman2017reverberation}. \\
\bottomrule
\end{tabularx}
\end{table}

\begin{table}[!b]
\centering
\caption{Timbre and tone-colour features: definitions, extraction library, and supporting reference.}
\label{tab:features-timbre}
\begin{tabularx}{\linewidth}{@{} K F L Y @{}}
\toprule
\# & Feature & Library & What it captures \\
\midrule
1 & \texttt{Spectral centroid}         & librosa~\cite{mcfee2015librosa}      & Brightness: spectral centre of mass~\cite{grey1977multidimensional}. \\
2 & \texttt{Spectral flatness}         & librosa~\cite{mcfee2015librosa}      & Tonal vs.\ noise-like character of the spectrum~\cite{peeters2004large}. \\
3 & \texttt{MFCC timbre fingerprint}  & librosa~\cite{mcfee2015librosa}      & MFCC-based timbre signature and how much it drifts~\cite{peeters2004large}. \\
4 & \texttt{Spectral rolloff}          & librosa~\cite{mcfee2015librosa}      & Frequency below which most energy sits: thin/bright vs.\ full~\cite{peeters2004large}. \\
5 & \texttt{Harmonic-\allowbreak to-\allowbreak noise ratio} & parselmouth\allowbreak\ \cite{jadoul2018introducing} & Harmonic energy vs.\ noise; overall cleanliness~\cite{hanson1997glottal}. \\
6 & \texttt{Spectral contrast}         & librosa~\cite{mcfee2015librosa}      & Peak-to-valley spread across bands: definition/clarity~\cite{peeters2004large}. \\
7 & \texttt{Instrument onset timbre}      & librosa~\cite{mcfee2015librosa}      & Timbral sharpness of note attacks~\cite{grey1977multidimensional}. \\
8 & \texttt{Breathiness H1-H2}          & parselmouth\allowbreak\ \cite{jadoul2018introducing} & H1--H2 measure of breathy vs.\ modal phonation~\cite{hanson1997glottal}. \\
\bottomrule
\end{tabularx}
\end{table}

\begin{table}[!b]
\centering
\caption{Vocal quality features: definitions, extraction library, and supporting reference.}
\label{tab:features-vocal}
\begin{tabularx}{\linewidth}{@{} K F L Y @{}}
\toprule
\# & Feature & Library & What it captures \\
\midrule
1 & \texttt{Formant frequencies} & parselmouth\allowbreak\ \cite{jadoul2018introducing} & F1/F2 formants and the resulting vowel space~\cite{petersonbarney1952}. \\
2 & \texttt{Jitter \& shimmer}      & parselmouth\allowbreak\ \cite{jadoul2018introducing} & Cycle-to-cycle pitch (jitter) and amplitude (shimmer) perturbation: vocal strain~\cite{farrus2007jitter}. \\
3 & \texttt{Vocal register}      & librosa~\cite{mcfee2015librosa} & Chest/head register classification and register breaks~\cite{lee2023registers}. \\
4 & \texttt{Diction clarity}     & Whisper~\cite{radford2023whisper} & Clarity/intelligibility of enunciated words~\cite{fineginsborg2014making}. \\
5 & \texttt{Breath support}      & librosa~\cite{mcfee2015librosa} & Sustained-note stability as a proxy for breath control~\cite{solomon2000respiratory}. \\
6 & \texttt{Lyrical syllable timing alignment}     & Whisper~\cite{radford2023whisper} & Timing and rate of syllables~\cite{ibrahim2017intelligibility}. \\
7 & \texttt{Vowel duration ratio}      & Whisper~\cite{radford2023whisper} & Length of vowel segments~\cite{roadabike2021voicesource}. \\
\bottomrule
\end{tabularx}
\end{table}

\paragraph{Structural segmentation.}
The system constructs a self-similarity matrix from each track's chroma features. It applies a Foote-style checkerboard kernel to obtain a novelty curve and uses its peaks as candidate section boundaries. Segments shorter than approximately eight seconds are merged with neighbouring segments. Rules based on position and repetition patterns assign the structural labels shown on the timelines. These rules provide navigation cues; they are not a trained classifier of musical form.

\paragraph{Comparison and interpretation.}
The browser sends the selected comparison scope to the comparison engine. The engine retrieves the relevant track or segment records and computes differences between corresponding feature values. It then sends these differences and a glossary of feature definitions to a self-hosted language model through vLLM's OpenAI-compatible endpoint. The model returns category summaries and an overall verdict. Schema-constrained decoding sets the response format, while a low sampling temperature limits variation in the generated text. The interface displays the returned interpretation with the feature comparison. Track features are reused when the scope changes; the comparison and model request use the newly selected pair.

\section{Method}

\subsection{Experiment Design}
\label{sec:experiment-design}
We conducted a researcher-facilitated, single-session, within-participants study to examine how musicians' comparative judgments changed after they used \oursystem{}. We staged the study as a before-and-after comparison rather than splitting participants into separate conditions. Asking participants to listen before seeing any analysis preserved an initial record of what they heard and where they located it; the subsequent system-use stage then allowed us to examine which judgments they retained, revised, withdrew, or added. This ordering also avoids treating the system's interpretation as the participant's starting point. Accordingly, the study evaluates changes following exposure to the complete system, rather than attributing those changes to the AI component alone.

\begin{figure}[t]
    \centering
    \includegraphics[width=1.0\textwidth]{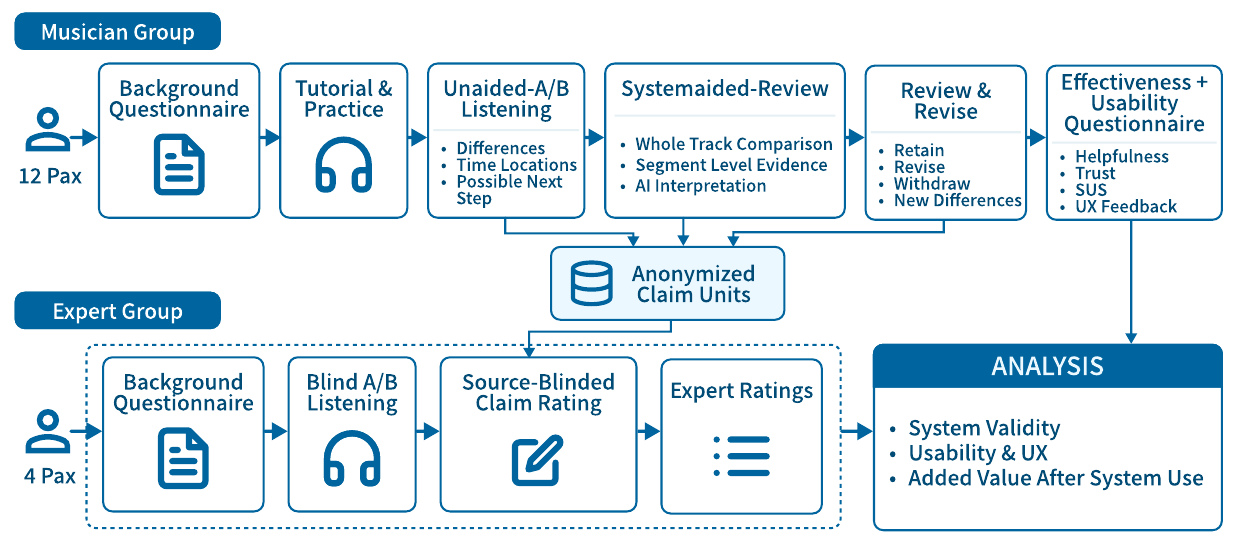}
    \caption{Experiment design. 12 musicians completed a background questionnaire, tutorial and practice task, unaided A/B listening, system-assisted comparison, judgment review and revision, and an effectiveness and usability questionnaire. 4 expert listeners completed blind A/B listening and source-blinded ratings of anonymized claim units, providing an expert perceptual reference for the analysis.}
    \Description{\descExperiment}
    \label{fig:experiment-design}
\end{figure}

After consent, participants used headphones and calibrated playback volume, completed a brief questionnaire about their musical background and reference-track experience, and received the same tutorial. They then practised the player, track switching, segment navigation, and evidence inspection on a separate song pair. Practice responses were excluded from analysis. Each participant subsequently completed one formal comparison using a predetermined target--reference pair supplied by the research team; the task did not require participants to upload their own music or make edits in a digital audio workstation.

\paragraph{Unaided listening.}
Participants first used a basic A/B player with no feature visualizations, system analyses, or AI output. They recorded the differences they heard, the relevant musical or production dimension and direction, an approximate time location where applicable, the audible basis for the judgment, and their confidence. They also noted possible next steps they would take if the target track were their own work. These initial responses were saved before the system-assisted stage and remained visible but uneditable as the pre-system record.

\paragraph{System-assisted comparison and revision.}
Participants next explored the complete \oursystem{} interface. They could inspect whole song comparisons, select and compare segments independently, review feature values and supporting evidence, and consider AI interpretations at either scope. The interaction did not prescribe a traversal order within the system. Participants then revisited every initial judgment and marked it as retained, revised, withdrawn, or still uncertain. They could also record newly noticed differences, their agreement or disagreement with the corresponding system view, and revised possible next steps. This stage treats disagreement as meaningful data: participants were instructed that AI output could be incomplete, inaccurate, or inappropriate for their creative judgment.

\paragraph{Post-task measures and logs.}
Participants completed a post-task questionnaire comprising system-effectiveness items, direct helpfulness and trust ratings, the System Usability Scale (SUS)~\cite{brooke1996sus}, system-specific interaction-clarity items, and open-ended feedback. We also logged time spent in the task, playback and track-switching actions, segment selections, and expansions of system insights or evidence. Together, these measures allow the study to relate reported usefulness and usability to how participants examined and acted on the available comparison information.

\paragraph{Expert perceptual reference.}
In a separate validation lane, 4 expert listeners first conducted blind A/B listening of the same materials without access to \oursystem{} or participant responses. The participants' claims from before and after using the system were then segmented into anonymized claim units and rated without revealing their source. The experts assessed whether each claim corresponded to an audible difference, whether its direction and temporal location were appropriate, and whether the claim was sufficiently specific and verifiable. We treat these assessments as an expert perceptual reference for interpreting agreement and disagreement, rather than as an absolute ground truth about the recordings.

For each track pair, the two assigned experts independently rated every anonymized claim using a three-level ordinal rubric. A claim received \emph{full support} when the audible difference, its stated direction, and its temporal location, where applicable, were supported as described. It received \emph{partial support} when the core difference was audible but one or more details, such as its direction, magnitude, wording, or temporal location, were not fully supported. It was rated \emph{unsupported} when the claimed difference was not audible or was inconsistent with what the expert heard. Before the formal evaluation, all experts received the same written rubric. Formal ratings were completed independently, and disagreements were retained rather than adjudicated. In the analysis, a claim was classified as fully supported by both experts only when both assigned the full-support rating; it was classified as at least partially supported by both when each expert assigned either full or partial support.

\subsection{Participants}
Participants were recruited through campus posters and online social media platforms. We recruited 12 musicians and 4 experts. The 12 musicians were randomly assigned to 4 target--reference track pairs, with 3 musicians per pair, forming four pair-specific groups. Each musician completed one comparison using the assigned pair. Experts took part in a separate evaluation of the claims collected from the task and the system; each expert was assigned two music pairs.

The musician group represented a range of performance, composition, music-related educational backgrounds, and hobbyist backgrounds, with reported music-related experience spanning from 1 year to more than 10 years.

The expert group reported at least three years of music-related experience, with backgrounds in composition, arranging, or performance. All 4 experts reported prior DAW use, and their responses indicated regular experience with several audio-analysis and music-production tools.

The study was approved by our university's Institutional Review Board. Participants provided informed consent before taking part. The musician task instructions specified compensation equivalent of 16 USD.

\subsection{Analysis Methods}
We analyzed the data descriptively. We matched pre- and post-use judgments on the same topic, excluded entries without an explicit follow-up or with a changed topic, and analyzed newly added judgments separately. We summarized participants’ review decisions and compared expert support before and after system use, reporting both full support and at least partial support from both experts. We examined participants’ evidence references and explicit disagreements to characterize how they used system interpretations, and compared confidence changes with changes in expert support for judgments with complete confidence ratings. Questionnaire responses were summarized using counts and SUS\cite{brooke1996sus} descriptive statistics, while open-ended feedback contextualized reported benefits and difficulties.

\section{Results}
\label{sec:results}

In this section, we begin by outlining the study data and the analytic sample. We then examine how participants’ comparative judgments and their expert support changed after system use, analyze how participants responded to the system’s interpretations, and finally report perceived benefits, usability, and difficulties.

\subsection{Results Overview}
\label{sec:results-5-1}

We collected task responses and post-study questionnaires from 12 participants who completed comparisons across four target–reference track pairs. The task data included 44 pre-use judgments, 39 post-use judgments, and 10 additional judgments reported after system use. After accounting for four unchanged judgments duplicated across the pre- and post-use records, the dataset contained 89 distinct judgments. Each judgment was evaluated by two experts, yielding 178 expert ratings. The primary before-and-after analysis comprised 38 topically matched judgment pairs. We excluded one post-use judgment because it addressed a different topic from its corresponding initial judgment. Initial judgments without an explicit post-use response were also excluded from the paired analysis rather than inferred to have been retained or withdrawn. Overall, participants found the system usable and supportive of comparison, but expert ratings showed no aggregate improvement in support for their existing judgments.

\subsection{Changes in Comparative Judgments and Expert Support}
\label{sec:results-5-2}
Expert support for participants’ existing judgments changed little after system use. Among the 38 matched pairs, 24 initial judgments (63.2\%) and 23 post-use judgments (60.5\%) were rated as supported by both experts. Most pairs remained in the same support category: 22 received full support at both stages, while 13 did not receive full support at either stage. One judgment gained full support after system use, and two lost it. When partial support was also included, 32 judgments (84.2\%) were at least partially supported by both experts at each stage.

After using the system, 7 participants also reported 10 judgments that they had not recorded during initial listening. Of these judgments, 9 received at least partial support from both experts: 4 were fully supported by both, 3 were fully supported by 1 and partially supported by the other, and 2 were partially supported by both. Only one was rated as unsupported by both experts. These findings show that the system contributed additional, expert-supported information to participants’ comparisons, even though support for their existing judgments remained largely unchanged.

\subsection{Responses to System Interpretations}
\label{sec:results-5-3}
Participants incorporated system information into their judgments in different ways. In our preliminary coding of the 38 matched judgment reviews, 17 evidence entries named only a panel, metric, or caption, whereas eight reported a metric direction, numerical value, or segmentation result. Naming system information did not necessarily establish how it supported the judgment. For example, G3-P03 retained a distinction between piano and xylophone-like timbres and cited a higher spectral centroid as evidence. Both experts supported the judgment, but one questioned whether the cited metric justified that distinction. Participants also challenged system interpretations: across six review entries, four participants explicitly expressed disagreement or identified missing evidence. These responses included rejecting an account of dynamic compression, questioning whether a 1.3 BPM difference explained a perceived tempo difference, and noting the absence of relevant harmonic or instrumental evidence. Such responses showed selective acceptance of system information, although disagreement with the system did not itself establish that the retained judgment was supported by experts.

Among 35 judgments with complete confidence ratings before and after system use, confidence increased for 15, remained unchanged for 17, and decreased for three. At the participant level, mean confidence increased for seven participants, remained unchanged for four, and decreased for one. However, changes in confidence did not consistently correspond to changes in expert support. Of the 15 judgments with increased confidence, seven retained full support from both experts, seven remained below that threshold, and one lost full support from one expert. For example, confidence in G1-P01’s modulation judgment increased from 75 to 96, while one expert continued to judge it unsupported and the other partially supported. These patterns indicate that increased confidence could accompany both supported and contested judgments.

\subsection{Perceived Benefits and Difficulties in Using the System}
\label{sec:results-5-4}

\begin{figure}[t]
  \centering
  \includegraphics[width=1.0\linewidth]{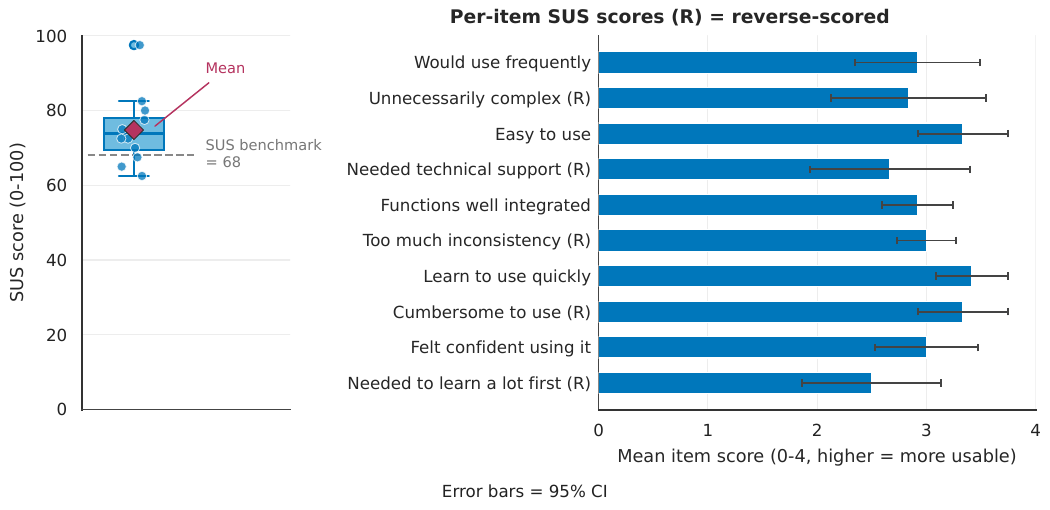}
  \caption{System Usability Scale. Per-item scores reverse-scored to a 0 to 4
  range, so a higher score is more usable for every item, including the
  negatively worded (R) items, $N = 12$.}
  \Description{\descSUS}
  \label{fig:sus}
\end{figure}

\begin{figure}[t]
  \centering
  \includegraphics[width=1.0\linewidth]{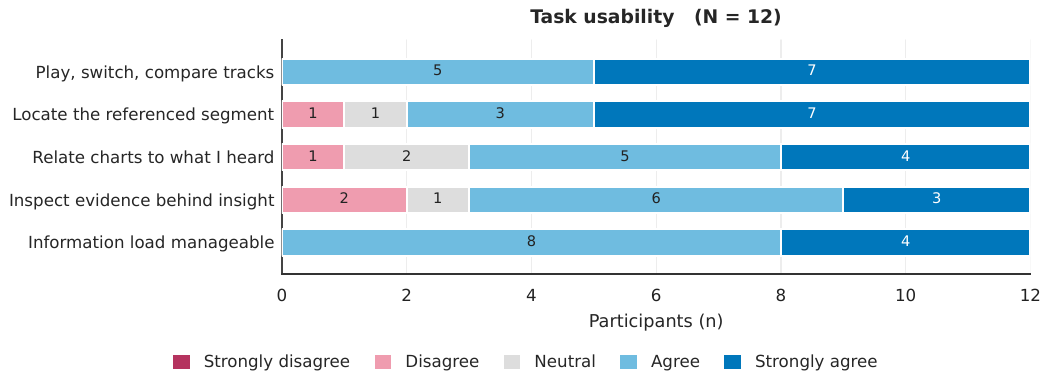}
  \caption{Task usability. Counts of agreement responses for five interaction
  items on a five-point scale (1 = strongly disagree), $N = 12$.}
  \Description{\descUsability}
  \label{fig:usability}
\end{figure}

\begin{figure}[t]
  \centering
  \includegraphics[width=1.0\linewidth]{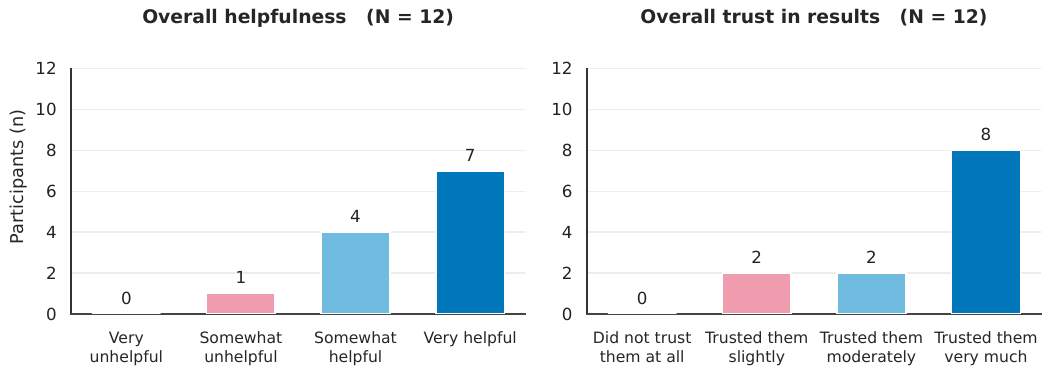}
  \caption{Helpfulness and trust. Counts for two single-item ratings: overall
  helpfulness for the comparison task and overall trust in the comparison
  results, $N = 12$.}
  \Description{\descHelpTrust}
  \label{fig:helptrust}
\end{figure}

Participants generally found the system easy to use and helpful for comparing tracks. Across 12 questionnaire responses, the mean System Usability Scale score was 74.79 (SD = 9.26, range = 62.5–97.5), above the commonly used average benchmark of 68~\cite{bangor2009determining}(Figure~\ref{fig:sus}). All participants agreed that playback and track switching were convenient and that the information load was manageable (Figure~\ref{fig:usability}). Eleven rated the system as somewhat or very helpful, whereas one rated it as somewhat unhelpful (Figure~\ref{fig:helptrust}).

\begin{figure}[t]
  \centering
  \includegraphics[width=1.0\linewidth]{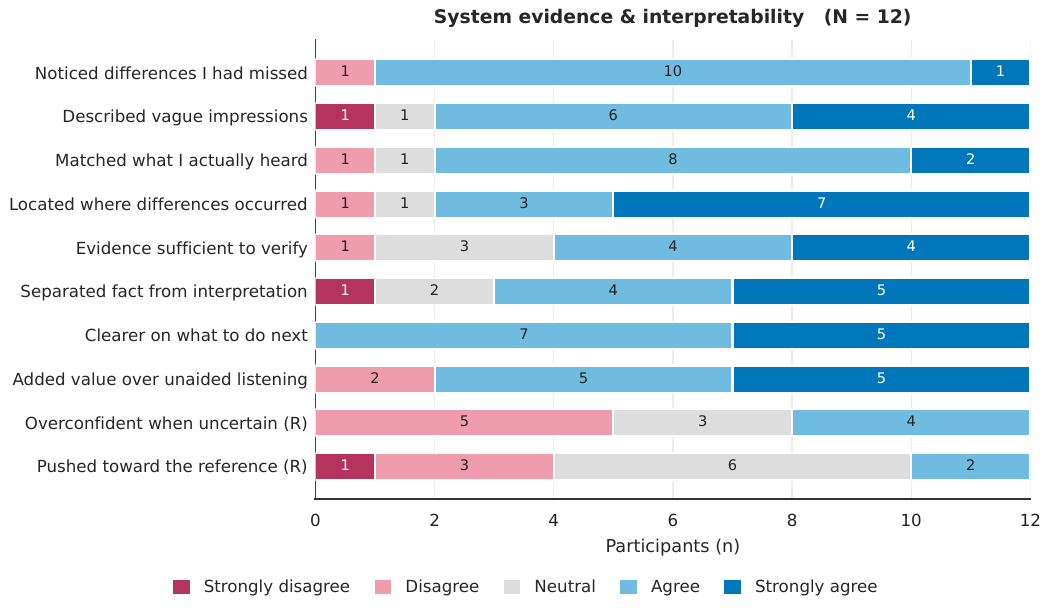}
  \caption{System evidence and interpretability. Stacked counts of agreement
  responses for ten items on a five-point scale (1 = strongly disagree),
  $N = 12$. Axis labels are shortened; full item wording is in the
  supplementary material. Items marked (R) are negatively worded, so for those
  a lower level of agreement is the favorable direction.}
  \Description{\descInterp}
  \label{fig:interp}
\end{figure}

Perceived benefits centered on making musical differences more explicit and identifying next steps (Figure~\ref{fig:interp}). Eleven participants agreed that the system helped them notice differences not recorded during initial listening; ten reported that it helped them describe vague impressions more specifically, and ten found the segment and time evidence helpful for locating differences. All twelve reported greater clarity about what to check, try, or leave unchanged next. Open-ended responses highlighted segmentation and time markers as aids to structural comparison and navigation, and A/B switching as a way to check AI statements through listening.

Understanding and verifying the analysis received less consistent support. Eight participants agreed that the evidence was sufficient to verify the system’s statements (Figure~\ref{fig:interp}), and nine could relate the charts to what they heard (Figure~\ref{fig:usability}). Open-ended responses described difficulties interpreting technical terms and charts, alongside requests for interpretations that better accounted for arrangement and instrumentation. One participant, for example, felt that describing a track as “less energetic” overlooked differences between its acoustic arrangement and the reference’s electronic percussion. Other concerns included overly general summaries and possible analysis errors.

Eight participants reported trusting the results very much (Figure~\ref{fig:helptrust}), although four agreed that the system sometimes made insufficiently supported claims too confidently (Figure~\ref{fig:interp}). Two also reported pressure to move the target toward the reference against their creative judgment. These findings indicate perceived support for comparison, alongside concerns about interpretability and creative autonomy.

\section{Discussion}
We read the results against the three questions that shaped the study. Participants liked the interaction, and they felt the system supported the comparison task. Even so, expert-rated judgment quality did not rise overall. These two results fit together rather than clash. The rest of this section explains why, and draws out what follows for the design of AI-assisted music comparison tools.

\subsection{Multi-scale comparison turns listening impressions into inspectable questions}
\label{sec:disc-multiscale}
The two modes serve different moments in the same task. Whole Song comparison gives an overall orientation. Selected Segments comparison lets a user go back to one spot and look at a difference more closely. Together they support a simple loop: start from a broad impression, drop down to local evidence, then return to the whole track. Participants reported this kind of use. All 12 said they were clearer about what to check, try, or leave unchanged (Figure~\ref{fig:interp}). In the open responses, several valued the automatic segmentation and the time markers, because they no longer had to keep the timing of one passage in mind while playing the other~\cite{nees2016auditorymemory}. Comparison across dimensions against the reference works the same way. The user picks what to hold up for comparison based on the question they are actually asking, not on a fixed alignment. Taken together, the design and the feedback point to one reading of what the system is for. Its main value may not be that it settles a comparison. It is that it helps a creator organize attention and decide what is worth a closer listen or a change worth trying.

\subsection{Perceived support without a measured gain in judgment quality}
\label{sec:disc-tension}
The main result is easy to state. Participants found the system helpful, but the experts did not rate their judgments as any better supported overall. Among the 38 matched judgment pairs, 24 initial judgments had full support from both experts, compared with 23 judgments after system use. These two results do not clash. Noticing more differences, describing them more clearly, and being right more often are three separate things. A judgment can become sharper and better placed in time without being any more likely to match what an expert hears.

The system did widen what participants paid attention to. Along with revising judgments they already had, they wrote down new differences they had missed on the first listen. This is worth taking seriously on its own. But it came with a cost. Of the 10 new judgments, 4 had full support from both experts, and 9 received at least partial support from both experts. Looking wider turned up more possible differences than listening alone, and more of them were weakly grounded. Participants also reported trouble with some terms, charts, and interpretations. That points to a gap between getting information and knowing how to use it. Some participants took up part of what the system said and set the rest aside. We do not read unchanged judgments, on their own, as proof that people kept their own creative view. The clearer sign of that is elsewhere. Most paired judgments were kept after using the system, fewer were revised, and a few were withdrawn. And only 2 of 12 participants said they felt pushed to move their track toward the reference against their own judgment (Figure~\ref{fig:interp}). That direct response is a better basis for the autonomy claim than the absence of change.

\subsection{The system described detail it could not ground}
\label{sec:disc-grounding}
Some of the expert disagreements were not close calls. They were confident descriptions of things that were not in the audio. Experts flagged a shuffle feel and short key changes they did not hear. They flagged emotional qualities they judged absent. In instrumental tracks, they flagged descriptions of a singing voice and its register. 2 participants ran into this directly, 1 saw the system report a vocal part in instrumental music, and another got a failed rhythm-intensity reading. 4 of the 12 agreed that the system sometimes stated uncertain things too confidently (Figure~\ref{fig:interp}). We read these cases as one pattern, not a handful of separate faults. The features set a limit on what the system can measure. They do not set a limit on how sure the written verdict sounds. So the verdict can add detail that sounds musical and plausible but that the measurement never supported. This points to a clear design change. Each sentence in the verdict should be tied to the feature or measurement it rests on, so the reader can see what backs it. And the system should stay quiet about dimensions it did not detect, rather than describing a voice in a track that has none~\cite{amershi2019guidelines}.

\subsection{More fluent judgments were not more accurate judgments}
\label{sec:disc-fluency}
Reading the before and after judgments side by side, the main change was in form, not in correctness. The later judgments were often longer and used more technical words. They cited tempo in beats per minute, stereo width, transients, and compression, where the earlier version had noted a plainer impression. This extra detail did not come with a matching rise in expert support, as the before and after numbers above show. The worry is that the system hands users a fluent technical vocabulary that makes a judgment sound more rigorous without making it more accurate. A claim can sound more polished and more sure than the impression it replaced and still rest on no firmer ground. Because it sounds better, the weak footing is harder to notice, both for the user and for anyone they explain it to later. This fits earlier findings that AI feedback can match human feedback in fluency and structure while falling behind on accuracy and on picking out what matters~\cite{steiss2024comparing}. A tool like this should show how strong the evidence is next to the wording, so that fluent writing is not taken as a sign of reliable content.

\subsection{Supporting evidence-informed judgment while preserving creative intent}
\label{sec:disc-intent}
The design lesson from the sections above is simple. For any given difference, the system should help the user do 3 things: find where it is, see what its interpretation rests on, and decide whether it matters for their own goals. In this framing the reference track is an anchor for understanding the work in progress. It is not a target the work has to be pushed toward. Most participants already treated it this way, with only 2 of 12 reporting any pressure to match the reference. The analysis should help a user decide which differences are worth acting on and which are worth keeping, instead of treating every difference as a fault. The expert results also carry a lesson for evaluation. Claims about things that can be measured directly, such as tempo, loudness, and key, tended to draw agreement. Claims about looser qualities, such as feel or emotion, drew more mixed responses, and here the two experts sometimes disagreed with each other. Musical judgment holds both kinds of content: attributes that can be checked, and readings that are aesthetic and depend on context. Expert agreement is strong evidence for the first kind. It cannot capture the full value of creative help. So evaluation of tools like this should keep weighing three things together: how reliable the judgment is, how good the supporting evidence is, and how useful the output is for a real creative decision.

\subsection{Pointing earned more trust than concluding}
\label{sec:disc-pointing}
The parts of the system people trusted most were the ones that helped them listen, not the ones that listened for them. Every participant could play, switch between, and compare the two tracks. Ten or more agreed the system helped them find where differences happened. In the open responses, the time and segment markers were among the most valued features. These features share one thing. They send the user to a place to listen, and they do not depend on the system being right about what a difference means. Navigation, section markers, and side-by-side playback are correct as long as they take the user to the right spot. A written verdict is only as good as its reading of the music, and that is where the trust concerns showed up. This suggests a design principle for tools of this kind. Features that help a person listen more efficiently earn trust more steadily than features that hand down an answer, because the first kind cannot be wrong the way the second kind can. A sensible split is to let the system be confident about where to look and more careful about what a difference finally means, and to leave that last call to the user. The trust data adds one caution. A very usable system can still be wrong. The participant who gave the system its highest usability score, and reported high trust, also reported the false vocal reading. A smooth interface should not be read as a sign of reliable output, and the two are worth reporting apart.

\section{Limitations and Future Work}
The expert evaluation has limits that come from its design. The experts were trained before rating, but the task still left room for individual interpretation. They may differ in how they read musical context, the wording of a participant's judgment, and whether the supporting evidence is enough. Expert ratings should therefore be read as assessments of support within our evaluation framework, not as final measures of correctness. A judgment that did not receive full support from both experts is not necessarily wrong.\par
On the feature side, not all features are equally accurate or equally useful. The pipeline reads 59 features from the audio files with standard tools, and some are more reliable than others. The study showed this directly. In one case the system reported a singing voice in instrumental music, and in another a rhythm-intensity reading failed. The features were also drawn from the literature rather than from formative work with producers, and participants questioned which of the reported numbers were useful and asked for closer attention to some dimensions. We did not measure which features matter most, so we cannot yet say which ones carry the comparison and which add little. Better and more targeted feature extraction is therefore our main line of future work. This includes checking each feature against reference annotations, making extraction more robust on hard cases such as instrumental tracks with no vocal, and studying with producers which features are worth showing in the first place.

\section{Conclusion}
We presented \oursystem{}, an AI-assisted system for reference listening that keeps the choice of what to compare with the user while supporting them in interpreting differences across whole tracks and selected segments. Our study with 12 musicians, complemented by source-blinded ratings from four expert listeners, showed that \oursystem{} helped participants surface additional, largely expert-supported differences and clarify what they might check, try, or leave unchanged. At the same time, expert support for existing judgments remained largely stable, and participants sometimes questioned or rejected the system’s interpretations. These findings suggest that the value of AI-assisted creative comparison lies not in replacing human judgment with more confident conclusions, but in making comparison more navigable, inspectable, and actionable. More broadly, we argue that systems supporting reference-based creative work should let users define comparison scope, expose the evidence behind interpretations, and treat AI outputs as resources for judgment rather than authoritative verdicts. AI can help point users toward what deserves attention; deciding what a difference means; and whether it matters; should remain with the user.

\section*{Acknowledgments}
This work has received support from MOE under grant number MOE-T2EP20124-0014, SUTD GAP-052 project, the National NaturalScience Foundation of China (No.62272409) and the China Scholarship Council program (CSCNo.202506320202).
\bibliographystyle{plain}
\bibliography{reference}
\FloatBarrier

\appendix

\end{document}